%% file: template.tex
\documentclass[preprint]{vgtc}               % preprint

\usepackage[utf8]{inputenc}

\ifpdf%                                % if we use pdflatex
  \pdfoutput=1\relax                   % create PDFs from pdfLaTeX
  \usepackage{graphicx}                % allow us to embed graphics files
  \DeclareGraphicsExtensions{.pdf,.png,.jpg,.jpeg} % for pdflatex we expect .pdf, .png, or .jpg files
\else%                                 % else we use pure latex
  \usepackage{graphicx}                % allow us to embed graphics files
  \DeclareGraphicsExtensions{.eps}     % for pure latex we expect eps files
\fi%

\graphicspath{{figures/}{pictures/}{images/}{./}} % where to search for the images

\usepackage{microtype}                 % use micro-typography (slightly more compact, better to read)
\PassOptionsToPackage{warn}{textcomp}  % to address font issues with \textrightarrow
\usepackage{textcomp}                  % use better special symbols
\usepackage{mathptmx}                  % use matching math font
\usepackage{times}                     % we use Times as the main font
\usepackage{cite}                      % needed to automatically sort the references
\usepackage{tabu}                      % only used for the table example
\usepackage{booktabs}                  % only used for the table example
\newcommand{\revision}[1]{#1}

\newcommand{\challengesection}[1]{\subsubsection*{#1}}

\onlineid{8301}

\vgtccategory{Research}

\ieeedoi{10.1109/BELIV51497.2020.00008}

\vgtcinsertpkg

\input{sections/00-title}

\input{sections/01-authors}

\teaser{
  \centering
  \includegraphics[width=\linewidth]{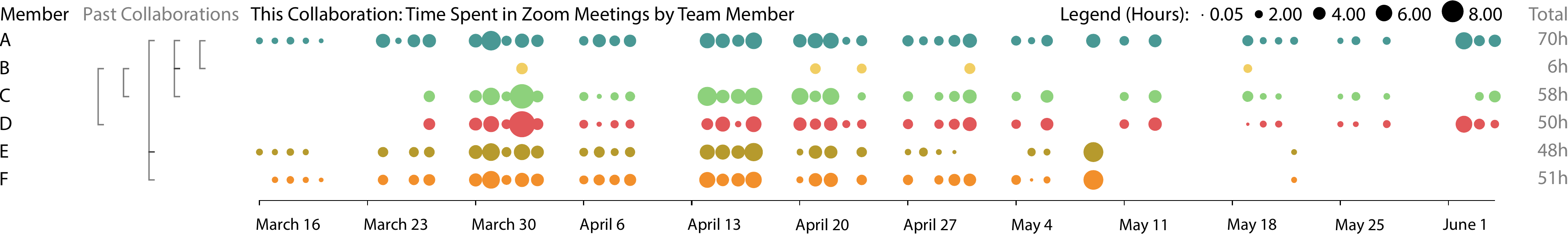}
  \caption{Overview of time spent video conferencing from March 16 to June 4 with concentrated synchronous design activities visible from late March to mid May; pivot points in the design process seen on April 2 when members C and D met for a prolonged design session using dual cameras as discussed in Section \ref{subsubsec:simulating-colocation}. See Table \ref{tab:overview-of-designers} for team roles. }
  \label{fig:teaser}
}

\input{sections/02-abstract}

\begin{document}

\input{sections/10-introduction}

\input{sections/20-related-work}

\input{sections/25-our-challenge}

\input{sections/30-our-design-process}

\input{sections/35-defining-our-needs-as-designers}
\input{sections/40-challenges}
\input{sections/50-discussion}

\input{sections/60-conclusion}


\acknowledgments{The authors would like to thank Jackie Yu, Neil Chulpongsatorn and the Centre for Health Informatics, Cumming School of Medicine at the University of Calgary for their contributions to
the process. This research was supported in part by the Natural
Sciences and Engineering Research Council of Canada (NSERC) and the European Union’s Horizon 2020 research and innovation programme under the Marie Sklodowska-Curie grant agreement No. 753816.}

\bibliographystyle{abbrv-doi}

\bibliography{template,remote-zotero.sk}
\end{document}

%% file: sections/00-title.tex
\title{Distributed Synchronous Visualization Design:\\Challenges and Strategies}

%% file: sections/01-authors.tex
\author{Tatiana Losev\thanks{e-mail: tatiana\_losev@sfu.ca}\\ %
        \scriptsize Simon Fraser University%
\and Sarah Storteboom\thanks{e-mail: sstorteboom@gmail.com}\\ %
     \scriptsize University of Calgary%
\and Sheelagh Carpendale\thanks{e-mail: sheelagh@sfu.ca}\\ %
        \scriptsize Simon Fraser University%
\and Søren Knudsen\thanks{e-mail: sknudsen@ucalgary.ca}\\ %
     \parbox{1.4in}{\scriptsize \centering University of Calgary}}

%% file: sections/02-abstract.tex
\abstract{We reflect on our experiences as designers of COVID-19 data visualizations working in a distributed synchronous design space during the pandemic. This is especially relevant as the pandemic posed new challenges to distributed collaboration amidst civic lockdown measures and an increased dependency on spatially distributed teamwork across almost all sectors. Working from home being ‘the new normal’, we explored potential solutions for collaborating and prototyping remotely from our own homes using the existing tools at our disposal. Since members of our cross-disciplinary team had different technical skills, we used a range of synchronous remote design tools and methods. We aimed to preserve the richness of co-located collaboration such as face-to-face physical presence, body gestures, facial expressions, and the making and sharing of physical artifacts. While meeting over Zoom, we sketched on paper and used digital collaboration tools, such as Miro and Google Docs. Using an auto-ethnographic approach, we articulate our challenges and strategies throughout the process, providing useful insights about synchronous distributed collaboration. 
} % end of abstract

%% file: sections/10-introduction.tex

\firstsection{Introduction}

\maketitle
In this paper, we describe and discuss our experiences of forming and working in a distributed visualization design team. While prior work has discussed visualization design processes~\cite{sedlmair2012design}, it tended to focus on how the visualization community works during face-to-face (co-located) design activities and processes for including users or domain experts. Less attention has been paid to how team collaboration works within teams of visualization designers, be they distributed or co-located.
% Would be great to cite here, but not sure we can find a clear example - thus, the phrasing "our reading".
%Building on the existing body of literature on collaboration in Human-Computer Interaction (HCI) and Computer Supported Collaborative Work (CSCW), w
We discuss our experiences of the differences between co-located and distributed visualization design work specific to our unique needs and experiences in light of an increased reliance on spatially distributed teamwork across almost all sectors. 

Our work was set in motion as part of a provincial response to the COVID-19 pandemic. We were a small team within a broader group of public health researchers who were providing data to government decision-makers during the pandemic. We formed our design team of six members (see Figure \ref{fig:teaser}, members A-F) to support the local health authorities' pandemic response in collaboration with our colleagues at the Centre for Health Informatics at the University of Calgary. As we were a newly acquainted multi-disciplinary team, we found it more beneficial to learn and to discuss the data together in a synchronous environment in the beginning of the project.  
We worked on the design of visualizations of provincial and national COVID-19 data for a public-facing website showing data visualizations of case and policy data. The intention with this site was to inform city- and province-level leaders to assist them in making sense of the status of the public health crisis and the associated data. 
While our team had visualization design expertise, our colleagues provided a wider set of skills and knowledge in fields such as public health, epidemiology, and data science. 
Due to the pandemic, we were unable to meet face-to-face. Instead and against our common work practice, all design team members worked from their home. This situation intensified challenges of distributed collaboration caused by distracting at-home work spaces and a sense of urgency amidst civic lockdown measures.  

Through an auto-ethnographic approach ~\cite{neustaedter2012autobiographical, bullough2001guidelines, duncan2004autoethnography, meyer2018reflection, meyer2020criteria}, we reflect on our experiences of distributed, primarily synchronous ideation and prototyping as a way to identify some challenges in our process of remote visualization design. %Auto-ethnographic research methodology is befitting because its primary evidence base is the unique lived experiences of the researcher(s) \cite{chang2013individual}. This framework was a useful way to discover relevant findings as they emerged via a process of iterative reflection upon our own data that we created throughout synchronous collaborative activities~---~assembling the team, frequent online meetings, data discovery and consultation with health experts, sketching, sharing and brainstorming ideas. The data that we used for our self-study are texts and artifacts from our design activities in Slack conversations, Google Docs, meeting notes, Miro board posts and sketches. 
Our main contribution is an articulation of challenges and strategies for dealing with these while doing distributed visualization design centered on the potential for doing activities, using technologies, and processes for communicating early and often, in order to reduce friction in visualization design teams.
% phase of remote collaboration in visualization design to enable a more fluid geographically distributed workflow.
 
% Due to strict isolation protocols mandated during the crisis, 
We relied on remote design tools and methods to actively design in real-time, often for two hours per day. We aimed to preserve the richness and diversity of co-located collaboration; specifically, the facilitation of face-to-face presence with the ability to view responsive body gestures, facial expressions, and the making and sharing of physical and digital artifacts.

%% file: sections/20-related-work.tex
\section{Related Work}

This work relates to literature on how to design visualizations; on how to support collaboration through visualizations; and on how to support design in technology-mitigated collaboration. 

\subsection{Designing Visualizations}
Making use of design methodology is widely recognized as important for creating useful and usable visualizations~\cite{munzner2014visualization, sedlmair2012design}.
This echoes the broader discussions about design~\cite{sedlmair2012design}, which span a wide array of concerns and advice about concrete design activities. Some examples that focus on co-located activities include the use of sketching~\cite{ buxton2007sketching, greenberg2011sketching, sturdee2019sketching, roberts2016sketching}; %brainstorming~\cite{};
prototyping with a range of media~\cite{snyder2003paper}, and the use of cards to support ideation~\cite{ideo}. 
A considerable amount of this design work is collaborative in nature ranging from the sharing of work in progress in design critiques~\cite{wolf2006dispelling}, through how people design together such as with co-creating artifacts~\cite{bannon2012design}, to ideas about how to involve people that are affected by a design, for example, through human-centered design~\cite{norman1986user} and participatory design~\cite{bannon2012design}. In addition, there are discussions about how to communicate designs, for example, in creating hand-off documents~\cite{maudet2017design} --- many of which people colloquially refer to as design (studio) praxis. 
In visualization, people have considered how to design for the intended audience such as specific experts or the general public (for example, design study methodology~\cite{sedlmair2012design, mccurdy2016action, meyer2020criteria}), and how to design with these people (for example, user-centered visualization design~\cite{koh2011developing, goodwin2013creative}). Also there are suggestions of potentially useful concrete design activities (for example: those based on constructive visualization~\cite{huron2016using}, those conducted as speculative design workshops~\cite{bressa2019sketching, knudsen2012exploratory}, and the use of pencil-and-paper based sketching of data visualizations~\cite{walny2015exploratory}). There is also advice on how to structure design processes, for example, how to teach visualization design to computer science students using five design sheet method~\cite{roberts2016sketching}), and how to more clearly communicate visualization designs (for example, in designers communicating visualization designs to developers as part of a hand-off process~\cite{walny2020data}). 
These are all extremely informative when at least some of the activities can happen co-located, however, they are less applicable for synchronous distributed design activities. In synchronous distributed design activities, they serve more as goals than as methods.  
These ideas will need to be re-interpreted in the context of synchronous distributed design work.

\subsection{Collaboration and Visualization}

There is a rich body of literature on collaboration in related fields such as human-computer interaction (HCI) and computer supported cooperative work (CSCW). The CSCW matrix (see Table~\ref{tab:cscw-matrix-in-vis-design}), which separates collaborative work along space time axes  is useful for understanding this area~\cite{grudin1994computersupported}. 
%We briefly discuss how this matrix relates to the visualization literature.
Two main areas of the CSCW matrix have been purposefully considered in visualization: co-located synchronous collaboration (Table~\ref{tab:cscw-matrix-in-vis-design}, top-left quadrant) and distributed asynchronous collaboration (Table \ref{tab:cscw-matrix-in-vis-design}, bottom-right quadrant). Although these two modes of collaboration have primarily been studied in isolation --- perhaps due to their different technological basis --- they share many challenges (see for example Isenberg et al.~\cite{isenberg2011collaborative}).
Research exploring co-located collaborative visualization includes  the use of tabletops for collaborative information access, for example, Scott et al.~\cite{scott2003system, scott2004territoriality}, the consideration of tabletop displays for collaborative browsing of hierarchical layouts of photographs~\cite{vernier2002visualization}, for analysis of scientific data~\cite{tobiasz2009lark}, and for exploration of book collections~\cite{thudt2012bohemian}.
More recently, people have considered large, high-resolution displays for supporting collaborative visual analytics (for example, Langner et al.~\cite{langner2019multiple} and Knudsen \& Hornbæk~\cite{knudsen2019pade}). 

Research on visualization for communicating across both space and time (asynchronous distributed) have led to ideas about democratizing visualization by making it accessible to all and to the increasing inclusion of visualizations as a way to communicate data in news media. Sense.us~\cite{heer2007voyagers} and ManyEyes~\cite{viegas2007manyeyes} introduced the collaborative possibilities for visualizations on the web. This type of research situates visualizations in a broader social and societal context. Based on these kinds of systems, Heer \& Agrawala~\cite{heer2008design} provide design considerations for collaborative visualization on the web more broadly. Later, work has shown that structuring the processes in collaborative asynchronous visual analysis can lead to increased analysis quality~~\cite{willett2011commentspace, willett2012strategies}, which might provide ideas for subsequent visualization designs~\cite{hullman2015content}.

While collaborative visualization-based analysis is distinct from collaborative visualization design, the CSCW matrix helps conceptualizing the space of synchronous distributed collaboration in relation to other collaborative contexts. There are also relevant similarities between collaborative visualization-based analysis and visualization design. For example, being able to point to a visualization or part of one is both important when designing and using visualizations~\cite{heer2008creation}. Supporting collaborative use of visualization is an important research direction. However, supporting visualization design in synchronous distributed settings has not yet been discussed. 

% Here, we put all the 'real' table contents. Then the table code below uses it in the right places.
% \newcommand{\sameplacesametime}[0]{\parbox[c]{2.75cm}{same place same time}}}
% \newcommand{\sameplacedifftime}[0]{\parbox[c]{2.75cm}{same place different time}}}
% \newcommand{\diffplacesametime}[0]{\parbox[c]{2.75cm}{different place same time}}}
% \newcommand{\diffplacedifftime}[0]{\parbox[c]{2.75cm}{different place different time}}}

\begin{table}[tb]
\centering
\caption{CSCW matrix for consideration in visualization design.}
% \begin{tabular}{@{}c|p{3cm}p{3cm}@{}}
\begin{tabular}{@{}p{1.85cm}|p{2.75cm}p{2.75cm}@{}}
\toprule 
& \parbox[c]{2.75cm}{Same time\\(synchronous)} & \parbox[c]{2.75cm}{Different time\\(asynchronous)} \\ \midrule

% \parbox[c]{2.0cm}{Same place\\(co-located)}
\raggedright Same place (co-located)
& same place same time & same place different time \\[0.5cm]
% \parbox[c]{2.0cm}{Different place\\(distributed)}
Different place (distributed)
& different place same time & different place different time \\ 
\bottomrule
\end{tabular}
\label{tab:cscw-matrix-in-vis-design}
\end{table}

\subsection{Collaborative Design of Visualizations}
While the visualization literature includes many discussions about design~\cite{beecham2020design,bigelow2014reflections,goodwin2013creative,koh2011developing,mccurdy2016action,mckenna2014design,mckenna2017worksheets,meyer2020criteria,munzner2014visualization,sedlmair2012design,walny2020data,hall2020design}, the focus is on the design processes rather than the collaborative process. The collaborations discussed  tend to focus on how visualization researchers collaborate in long-term projects with domain experts. For example, while the term ``collaboration'' (and related forms) is used 34 times in the design studies paper~\cite{sedlmair2012design}, it is only used a single time in the section that discusses the ``core phase'' of the design study methodology. Discussions about collaborative design thinking seem to be missing. 

Similarly, in CSCW literature, while there are discussions about many different types of work, the focus has been on distributed asynchronous and co-located synchronous. We are interested in distributed synchronous design activities.   

\subsection{Collaborative Design Processes}
\revision{There is CSCW literature about collaborative design processes (for some examples see~\cite{tausch2015thinking, fischer2004social, hilliges2007designing, yu2011cooks, arias2000transcending, bratteteig2012spaces, obendorf2009inter}). However, the focus is still about collaborative design when co-location is part of the design process. 
For example, some have explored technology mitigated collaborations through tabletop display tools~\cite{hilliges2007designing, bratteteig2012spaces}, and through investigating the impact of technology based feedback about the group creative design process~\cite{tausch2015thinking}. One suggestion is to explore the use of the crowd in design processes\cite{yu2011cooks}.
Visualization has been used to provide feedback on speaking times and speaking turns during collaboration~\cite{ bergstrom2007conversation}.}

\revision{There is some exploration of the space we are interested in – the space of how to re-kindle the benefits of co-located collaboration in a technology-supported, synchronous distributed situation. Arias et al.~\cite{arias2000transcending} start by acknowledging the complex design problems often require group solutions and articulates needed design support for urban design problems. Both Fischer~\cite{fischer2004social} and Obendorf et al.~\cite{obendorf2009inter} consider the complexity of team-based design needs where teams must cope with difference in time, space, and knowledge and make a call for deeper exploration of the needs of these types of design teams. Our work, which describes our experiences of the challenges of collaborative synchronous, distributed technology-mediated design, and a range of strategies for dealing with these challenges, contributes to this larger call for research.}

%% file: sections/25-our-challenge.tex
\section{Synchronous Distributed Visualization Design}

Amid the COVID-19 lockdown, we were faced with the urgent challenge to design a useful COVID-19 visualization. We were confronted with the reality that our familiar co-located team-based collaboration design approaches could not be directly applied in our enforced distributed but synchronous realm.
While this was a challenge in many ways, we managed to reach an effective design process. Through the use of a reflective auto-ethnographic research approach, we have obtained a deeper understanding of these challenges.
% Before discussing these, w
We first describe our research approach. 

\subsection{Methodology: Reflective Auto-Ethnography}
\revision{We conducted a team-based self-study for this project by combining an auto-ethnographic research approach~\cite{chang2013individual} with hermeneutic phenomenology~\cite{van2016researching}. Auto-ethnographic research is befitting because its evidence base is the direct narrative of the people involved~\cite{chang2013individual}. Hermeneutic phenomenology complements this as it studies the meanings of lived experience through self-reflection, writing, and discussion~\cite{van2016researching,laverty2003hermeneutic}.
Together, these approaches deepened our understanding of our experience as a team of designers creating visualizations together in a synchronous distributed context. Though more commonly seen in the social sciences, we benefit from these qualitative research approaches. They enable us to focus on our unique experiences of designing visualizations in a synchronous distributed setting through the discovery of themes that occurred in our non-linear collaborative practice. }

% Provided that data visualization methods and processes are often not linear and context specific, o

% ur research findings enrich a shared understanding of diverse socio-cultural dimensions. The themes we extracted from our account offer an example to the members of the viz community who may identify with and acknowledge similar experiences that they too had, which serve to amplify an experiential evidence base to inform future visualization research and methodology in this domain.}

\subsection{Demographics: Our Team Members}
\revision{Our group of designers worked together, as a team, for the first time, though some team members (AB, AEF, ABC, BC, and BD --- also shown in Figure \ref{fig:teaser}) had worked together previously on separate projects. The team dynamics and the social setting of the lockdown were novel to the whole team. The domain specific data and the needs of the project had yet to be learned. Thus, the team needed to get acquainted with one another as well as the data. We found that spending more time together in a synchronous setting facilitated an immediate peer-to-peer learning, and improved our communication --– in our experience, this was the most suitable way for us to connect, to build personal relationships in our group, and to improve group synergy. Initially, team members A, E, and F discussed the project for about two weeks, followed by members A and B discussing possible additional members to balance skills in data visualization, design, public health, and programming (see Table 2). Team member A assembled the team and called the first team meeting.}

\subsection{Our Process, Data, and Analysis}
\revision{Following an auto-ethnographic approach, the data is both a result of our process and fueled it; therefore, we discuss them as intertwined. We looked closely into the process of our visualization project by analyzing our own experiential data and project artifacts~\cite{chang2013individual,creswell2016qualitative,laverty2003hermeneutic}. In auto-ethnography we are both the participants and the researchers --- through a self-reflexive process~\cite{chang2013individual,{van2016researching}}, we examine our shared experience as a multi-disciplinary team of designers working towards gaining a deeper understanding of our experience. Doing so, we questioned: \emph{how has distributed collaboration shaped our experience of synchronous design processes?}}

%% need text to go with this
% As a first and important step, we acknowledged our relationships that may contribute to personal biases relevant to the project and detailed our previous experience of designing visualization and how it impacted our contributions to the project

\revision{Our process was as follows: We continuously collected process data and members of the team kept regular day-by-day written team notes; the team's visualization sketches were collectively stored on a Miro board; 
%(for an example see Figure~\ref{sketch.fig}). 
our design brief on Google Docs and the Slack history texts were gathered and reviewed; we created a visual timeline from our Slack history; we formulated questions to guide our inquiry and reflection based on our collected data and previous personal experiences relevant to visualization design and health communications; we characterized, analyzed and reflected on our texts, discussed the texts, and generated new texts;
we held reflective discussions and documented this in our notes in Google Docs; themes that emerged in our text and from our reflective dialogue were grouped; through our reflections, we interpreted our documented themes and continued to write these reflections on our account; we corroborated collective written experiences with one another through further discussion to validate the findings. Lastly, we wrote this paper by detailing our experiences and findings by repeatedly going through the steps above.}

% \begin{itemize} 
% %[noitemsep]
% \item We collected our data – regular day-by-day written team notes, team's visualization sketches, design brief, and Slack history texts. We created a visual timeline from our Slack history
% \item We acknowledged our relationships that contributed to personal biases relevant to the project
% \item We wrote questions to guide our study and reflection based on our collected data and previous personal experiences relevant to visualization work and health
% \item We acknowledged and detailed our previous experience of designing visualization and how it impacted our contributions to the project
% \item We characterized, analyzed and reflected on our texts, discussed the texts, and generated new texts
% \item We held reflective discussions and documented the notes in Google Docs; themes that emerged in our text and from our reflective dialogue were grouped
% \item We interpreted our documented themes and continued to write our account
% \item We corroborated collective written experiences with one another through further discussion to validate the findings
% \item We wrote this paper detailing our experience and findings after cycling through the steps above
% \end{itemize}

\begin{table}[tb]
\centering
\caption{Overview of our team.}
\begin{tabular}{@{}llp{4.5cm}@{}}
\toprule
Team member & Role         & Expertise in use                                               \\ \midrule
A       & PI           & Visualization, design, programming, management, communication       \\
B       & PI           & Visualization, design                                                  \\
C       & Design lead  & Visualization, design                                           \\
D       & Designer     & Public health, design                                         \\
E       & Intern       & Visualization, programming                                    \\
F       & Intern       & Visualization, programming                                    \\ \bottomrule
\end{tabular}
\label{tab2}
% I would like to make a better version of this for camera ready. I imagine you could show a column for each expertise, and then have a vertical label across the column that describes the skill
\label{tab:overview-of-designers}
\end{table}

 \subsection{Our Context}
March 13, 2020, two days before a local state of emergency was announced, the office of the mayor requested help from the Centre for Health Informatics to get a better sense of non-clinical interventions of COVID-19 locally, nationally, and internationally. At that time there was a generalized sense of fear and widespread sensationalism that was broadcast via a multitude of media. A team of 37 members of researchers in health and data sciences from the centre assembled on Slack to create a ``COVID-19 Working Group'' determined to research COVID-19 data in response to the urgent call for information. The hope was to help by contributing critical information to aid informed decision-making about managing the pandemic. Working under an immense sense of urgency, the team collected cumulative and daily case numbers, researched global COVID-19 policies, and worked on epidemiological disease models that informed the status of the pandemic on a local and national scale. This research was used by municipal and provincial policy-makers. 
Two team members promptly responded by using open COVID-19 data and a web charting library to, within a few days, assemble a website that tracked the changing local and provincial COVID-19 data. This website became known within the team as the ``COVID-19 Tracker'' ~\cite{COVIDTracker}. During the first 8 weeks, the site garnered 15,000 page views.% it would be better to give dates 'between xx and zz' - the pandemic is not necessarily over. 
% @misc{COVIDTracker,
%   author = {},
%   title = {{THE COVID-19 RESPONSE
%   How is Canada doing and what should we be doing next?
%   Centre for Health Informatics, Cumming School of Medicine, University of Calgary}},
%   year = 2020,
%   url = {https://www.chi-csm.ca/},
%   urldate = {2020-08-07}
% }

While delighted about the speed of this action, the need for a more carefully designed response was noted and a different team member assembled a smaller design team. It is this smaller group that is our design team and it is our actions in this team that we focus on. Our initial conversations in the design team were via a specifically formed Slack channel.
Our design team's dedicated channel decreased the amount of notifications to members of the COVID-19 Working Group and helped focus our discussions. 
%why
%We transitioned into a smaller design team %smaller than the 8 member?/
Additionally, our design team  started to meet using video conferencing.
We discussed COVID-19 design issues to better understand the impact of design on a broader cross-section of people.
% , not just team members but also more generally the public. %why
Our design team delved into the intricacies of the available public COVID-19 data with the goal of designing data visualizations that would support a broad cross-section of the population, which the working group had identified as important: provincial and municipal decision makers, public health officers, as well as the general public. %] %do we have an audience?
%understand the meaning of vast information about COVID-19. 
We met frequently --- often several hours a day and still met more than once a week with the COVID-19 Working Group to align with the project needs and direction in response to the status of the pandemic. %not clear

%% file: sections/30-our-design-process.tex
\section{Describing Our Design Process Experiences}

In this section we describe how we experienced the activities that we engaged in as the design team. In keeping with our auto-ethnographic and phenomenological methodologies, rigor in this report means staying true to the reality of our experiences in our team via detailed descriptions and iterative reflections on our texts and artifacts~\cite{laverty2003hermeneutic,creswell2016qualitative}.
In starting our distributed design process, we consulted with the literature. However, we discovered only limited advice on how to organize collaborative synchronous distributed visualization design processes. While we considered our readings about visualization design and distributed design in CSCW, we largely relied on our own, largely face-to-face, experience of prior design processes in visualization design and beyond. Here, we describe how we experienced synchronous distributed visualization design. By articulating challenges and strategies, we discuss the factors that arose in our experiences that may prove useful to consider when doing synchronous distributed visualization design.

\subsection{Establishing Meetings and Technologies}

The design team met to discuss and to sketch together for two hours a day, five days a week and attended half hour meetings with the broader team a few times per week. We mostly used Zoom~\footnote{https://zoom.com/} for meetings, Slack~\footnote{https://slack.com/} for short asynchronous communications, Google Docs~\footnote{https://docs.google.com/} for written notes and records, and Miro Board~\footnote{https://miro.com/} for collaborative design. 

Daily 2-hour meetings over Zoom brought the team together and allowed for developing an awareness of each other and established a social dynamic and rapport among the team members. The meetings were a time to convene and establish project expectations, sketch, and design together. Most of us had device cameras positioned to show our face during the Zoom meetings. The team also participated in the larger Zoom meetings with the COVID-19 Working Group to gain feedback on our visualization ideas and sketches, and to hear of new developments or requests from senior leadership. A Slack channel was the main hub to set up meeting times, to share reading and video materials about COVID-19 data and visualization design, and to inform each other of online events such as webinars. We also used the Slack channel to post our design ideas. Our visualization design team made  use of a collaborative digital whiteboard (Miro). This provided digital space for the group to post sketches, PDF’s, and virtual sticky notes during design meetings. Meeting notes and a design brief were created and stored in Google Docs and were used concurrently during our team meetings with one team member taking notes. After five weeks of daily Zoom meetings, our team reduced meeting times over a collective sense that meetings needed to be more directed and convergent — the ideation phase was wrapping up and the team was keen to implement the design.

\subsection{Defining a Purpose and Setting a Direction}
% \cite{segel2010narrative}
% \cite{hullman2011visualization}

% In our design process, we considered literature on using storytelling elements in visualization [cite]. For example, we considered the Martini Glass structure in order to scaffold the complexity of the data and to create cohesion between COVID-19 charts to deepen an understanding of the complexity at hand. We also considered critical discussions of visualization [cite]. For example, while we were interested in helping people understand relationships between otherwise disparate datasets such as case numbers and policies, we recognized that separating certain data from the chart was critical to ensure that correlation was not assumed by the viewers such as overlaying policy data over case numbers could be misconstrued as a causal relationship.
%  In starting our distributed design process, we consulted with the literature outlined above. However, this body of literature offers limited advice for how to organize collaborative distributed visualization design processes. Instead, while considering our readings about visualization design and CSCW distributed design, we drew from our own previous experience -- largely face-to-face -- design processes in visualization design and beyond. In the following, we describe our experiences with doing distributed visualization design and then, in the next section, we offer a discussion of factors to consider through an articulation of challenges and opportunities. 

To reach a shared understanding of the project expectations, including timelines and target audiences, we collaboratively authored a design brief. This document directed some of our discussions as we considered the purpose of the site, our audience, and their familiarity with the data. The ideation process began with a distributed face-to-face critique of the COVID-19 visualizations that were already available on our site and across all of the provincial sites in Canada. 
% In addition, we identified misleading COVID-19 data visualizations on a local public health site and were able to provide feedback to its creators and remedy the situation. 
Screenshots of the various visualizations were compiled on a Miro board along with suggestions for improvements. This in turn led to ideas for improvements to the COVID Tracker website. The activity of critiquing visual elements of other visualizations was a beneficial learning exercise that focused our sketching sessions and informed our design choices. This activity enabled an engaged exploration of the data, of the end-user audience, and the purpose and messaging of the visualizations. Importantly, this activity presented the complexity of the data and enabled us to identify further questions and consultations necessary to validate our findings. While we were interested in helping people understand relationships between otherwise disparate data sets, such as case numbers and policies, we recognized that separating certain aspects of the data was critical so as to not suggest causation where correlations might exist. For example, it became clear that juxtaposing policy data with case numbers could be misconstrued as a causal relationship. 
% We subsequently validated this concern with a researcher from our COVID-19 Working Group. 

\subsection{Regular Sketching Sessions}
The goal of our synchronous online meetings during the ideation phase was to generate many ideas and sketches, while gaining an understanding of COVID-19 data. We spent a lot of time considering and exploring the data; understanding testing rates, positive case numbers, hospitalization cases, and disease transmission along with policies and correlations. We formed questions through repeated discussions, which, in consultation with members of the COVID-19 Working Group, provided a rich method for developing an in-depth understanding of the data and issues of interest. 

Sketching was a valuable activity that helped us to think through concepts, envision a story, and share ideas. Sketching enabled our team to see the data and gain a shared sense of our individual perspectives. During the meetings, and while apart, we sketched on paper and tablets. The sketches were the main artifacts that we each created and showed to each other either through presenting our physical sketch to the device camera or posting it onto our Miro board. The sketches served as the foundation for our discussions.

The design meeting notes and artifacts were stored, categorized by date, and accessible to the team. The cache of sketches along with inspiration clippings and meeting notes proved very useful. We were able to refer to previously posted resources and sketches during design meetings, enhancing our ability to recall our previous work and build on it. However, we found the lag time when posting our paper sketches onto our virtual whiteboard to be challenging. We dealt with this by holding sketches up to the camera, but they were not easily referenced until they were scanned and added to the board.

%as references to our ideas during the ideation phase.
% Showing sketches over teleconferencing does not work with virtual backgrounds. %can you say more about this?

\subsection{Software Prototyping}
Illustrator versions of our design were produced by the lead designer so we could view a pixel-perfect design. After several iterations, the Illustrator file was handed off to three team members who implemented the design using D3~\cite{bostock2011d3}. Several iterations of the software prototype were critiqued during collaborative design team sessions and subsequently tweaked. The design phase for the data visualization continued through implementation as updating data sources presents new challenges to be solved. When prototypes were polished and reflected live data sources, they were presented in a Zoom meeting to the COVID-19 Working Group for feedback.

%% file: sections/40-challenges.tex
\section{Emerging Factors in Our Virtual Visualization Design}
Our purpose was to think critically about how to communicate the complexity of the pandemic and the data while ensuring that our visualization would not be misinforming. Our process, however, started by identifying the missing factors in our distributed design situation. The familiar lab environment facilitated serendipity, natural discussion, and a tangible sense of togetherness that allowed for ideas to spontaneously emerge. In order to support the process of ideation and data discovery in our distributed environment, we wished to collaborate via real-time sketching and discussion akin to a collocated design environment such as a lab. We searched for useful tools and materials to enable the team members to communicate ideas about interactions for the design and to share them, as we normally would, ``in-person''. 

We note that being distributed forced us to be upfront about the process. This in turn supported later reflection because our distributed work had been logged through the various tools used in the design process. This design experience was distinct as a result of distributed collaboration under the time sensitive demands of a public health emergency. Furthermore, remoteness posed a perceived risk of misunderstanding and miscommunication, more so given a flux of ever-changing public health data that decision makers were relying upon. This design experience elucidated several factors for re-consideration. Notably, these factors were initially experienced as challenges but sometimes, through working with these challenges, we also noted potential strategies and opportunities. 
%While doing distributed visualization design posed challenging as we have described above, this mode of working also provided opportunities. Here, we outline 
% We recognize that some of the experiences we describe seem to as easy, or potentially easier than co-located design work. 

%\fix{x} \textbf{
In the following, we describe eleven factors that emerged from an auto-ethnographic exploration of our distributed visualization design process. While the design factors encompass a wide array of considerations, we recognize that there are more opportunities and strategies possible through re-purposing current software and hardware tools. 
% We just mention the ones that we tried.
%\challengesection{C0. Prototypical challenge}
%\textit{Briefly (2-4 lines), What is the challenge}
%Longer explanation of why this challenge is important to be aware of.
%For example, \ldots
%Reflecting on the challenge
%(possibilities/or what we did to solve it).
%\challengesection{C1. Barrier to ‘face-to-face’ communication}
 \subsection{Pandemic Challenges}
 
%Our first challenge is specific to the situation during the COVID-19 pandemic lockdown. 
First and unsurprisingly, the pandemic lockdown led to challenges. These are important to recognise as they heavily influenced our ability to carry out work.
 
 \challengesection{C1. Negotiating Time and Resources}
\textit {Coordinating time and access to physical space for some team members was a challenge amidst the social lockdown. This challenge is important because the pandemic brought forward some social constraints to distributed work from home such as interruptions in each member’s home environment, lack of physical working space, and faulty hardware that played a role in how the team managed to design together remotely.} 
For example, some collaborators were forced to leave meetings in order to take care of their children and at times their children showed themselves to the camera during design meetings. There were instances when team members had to move their device to a different room during meetings because their small shared living spaces were prioritized based on homeschooling needs of children or for other family members who were working from home. We contended with web cams that did not function, so some team members were not visible during camera-to-camera meetings. Purchasing web cameras at the time was difficult due to a high demand in the market with increased remote work. As a team, we resolved to continue meetings during these circumstances. Audio was muted for a moment when children interrupted our online meetings and we waited until family disruption ended. We persisted with meetings though some members were not visible, interacting only via audio.
Additionally, this work provided an opportunity to ``be with'' other people and to ``contribute'' in a way that was a cathartic exercise for several team members in the midst of this pandemic.

 \section{Communication Breakdowns}
 
 This group of challenges focus on how our adaptation to distributed collaborations caused several types of communication challenges — some of which we learned how to mitigate.
 
\challengesection{C2. Establishing Team Cohesion}
\textit {The sense of uncertainty that arose steeply during the initial COVID-19 lockdown, along with our own perceived challenges of remote collaboration, posed risks of miscommunication and possible difficulties in establishing a spirit of team trust and cohesion.}
This was a challenge we were particularly aware of, because, while most of us were in the same geographical region, we designed synchronously from our homes. Our intention was to recreate the familiar ``face-to-face'' co-located communication flow even though we were working in a distributed virtual setting.
We chose frequent synchronous virtual collaboration as a way to 1) align our ideas and learn 2) mitigate a sense of uncertainty during a pandemic with frequent feedback 3) establish team cohesion and rapport. 
This virtual space was a new reality for many collaborators.
% due to the pandemic. 
Seeing one another brought a sense of togetherness and co-presence that was conducive to successful and natural synchronous teamwork. Notably, it was possible for our virtual interaction to achieve a similar communication workflow with the team’s facial expressions and gestures visible through our device cameras. Human communication and connections are an interplay of nuanced cultural mannerisms that may not be visible or as obvious through a webcam. Nonetheless, seeing each other through webcams and computer screens established a collective awareness and a sense of attendance within the team though we were not co-located. However, eye-to-eye contact differed from being co-located. We tended to look at each other's faces on the screen and not directly into our cameras, often seeming as though eye contact was averted during conversations. We found that group etiquette naturally developed in our online work space such as muting the microphone during excessive background noise, putting up a hand to speak, and waving goodbye into the camera. 
A couple of the team members chose to keep their cameras off during all online meetings but remained audible, which created a barrier to gaining an understanding of their affect and their level of engagement because gestures, facial expressions, and eye contact were not visible. We found that members who had their cameras turned on often dominated and contributed more to the discussion and invested more time throughout the project.
% \challengesection{C5. Creating a Team Working Space}

  \challengesection{C3. Understanding Team Members' Progress}
\textit {The challenge was to determine the progress of individuals on the team so that we could move forward with project goals. It was important to understand where team members were in their work to ensure that we were completing our goals.}
The team was rapidly assembled; some members were volunteers. They were unfamiliar with each other’s skills and were unsure if it was okay to ask others given the circumstances. Some work needed to be done asynchronously and, from time to time, team members were not able to follow through with tasks.
For example, we decided to assign tasks such as to scope nationwide provincial sites for visualizations of COVID-19 data and compile the findings in an Excel file, while another member was reviewing other information outside of our team meetings.
It was also valuable to prepare some information before our design team meetings to bolster our discussions about visualization design. To ensure that everyone was aware of the status of each task, team member A would check on the progress via Slack, which was visible to the whole team. This method was useful to ensure transparency and clear expectations for individual task completion within our design process. 
Some members, notably team members C and D often self-assigned tasks while other members relied upon A for task assignment and review.
 
 \challengesection{C4. Diversity of Software and Sketching Expertise}
\textit {There was considerable variation in levels of skill when sketching ideas or visualizing data, meaning that some of the team members struggled more than others with design activities and required guidance.} 
As a solution to some of these differences, the lead visualization designer mentored willing team members who were less experienced in designing data visualizations. For example, we held optional group sketching sessions via Zoom as a way to learn and to practice sketching apart from our team meetings. To bypass software knowledge gaps, we photographed physical paper sketches and shared them on the Miro board because everyone was able to use their device cameras comfortably. However, photographing our sketches stifled the natural flow of sketching and sharing during meetings, so we reverted to sketching and presenting our sketches through the camera. This method led to more productive discussions and more sketching to take place during the meetings.
We also saw a benefit in the virtual environment because sketches could be easily seen by everyone at the same time provided the camera was set up in such a way that the sketch was well lit and in focus. We would often scan sketches using a phone app or transfer them to the digital whiteboard, where we could continue to review the sketches.
After reviewing the sketches and discussing their features we decided on the best sketch to prototype. 
We also found screen sharing to be a benefit during the polishing stage of a working prototype. While designers might want to adjust some of the positioning and style of elements in a working prototype, they may not have a development space set up or the skills to make those adjustments. The programming team members utilized screen sharing to collaboratively make those adjustments in real time with the designers.
 
% \begin{figure*}[tb!]
%   \centering
%   \includegraphics[width=\textwidth]{figures/inspiration-cropped2.png}
%   \caption{An excerpt of our Miro board. In this area of the board, we compiled COVID-19 visualisations for shared inspiration, data understanding, and critique. We continually referred back to this area of the board during our design sessions.}%\soren{This can't be empty. Also, we need higher resolution}.}
%   \label{fig:inspiration-board}
% \end{figure*}
%  \subsection{Visualization Design Process}
  
 % This section discusses factors that arose throughout our design process.
  
\challengesection{C5. Understanding and Sharing Data}
\textit {Learning about the project needs and the health data was foundational to our visualization design, so we spent considerable time reviewing and discussing policy data, testing data, health care system capacity, social determinants of health, studying COVID-19 visualizations across Canadian jurisdictions, and in regular consultations with the COVID-19 Working Group in a remote setting.} 
As COVID-19 data was being visualized broadly across the world, we knew how the data was typically shown, so we considered the clarity and lack of clarity in existing visualizations. Therefore, we looked for new ways to visualize the same data that would not perpetuate the same issues. Information gathering was crucial to our design process because it enabled us to identify relationships and assumptions within the data as we sketched data visualizations in the early design stage. Notably, this process provided a fruitful distributed design environment with a serendipitous sense of data discovery. This collaborative online knowledge seeking provided a generative online learning space. Additionally, this activity allowed us to deepen our collective understanding of the data and cultivated team rapport.
This is an area where we identified a clear benefit to working in a distributed environment. We were able to work independently and then quickly share our findings when identifying something interesting. For example, we considered critical discussions of visualization, such as the understanding that positive cases were indicative of testing capacity. We followed this insight by brainstorming ways to visualize this relationship in a more clear manner as part of a larger visualization product. 
 % fill if possible ~\cite{} while reviewing the data with a sense of urgency; an extended amount of time was spent exploring the data and communication to the audiences with a heightened sense of risk if the data were misrepresented in our design. The activity of critiquing visual elements of other visualizations was a beneficial learning exercise that sparked productive questions and rich discussions.
\challengesection{C6. Sharing and Acquiring Knowledge}
\textit{It was challenging to share and introduce knowledge and background skills to other team members. Likewise, it was challenging to acquire new understanding from other team members.}
Distributed modes of collaboration introduce more friction in sharing knowledge. It is more difficult to find and share resources and point to specific parts of a resource as it is prone to error and requires time. This is particularly relevant in a visualization context. In contrast, in co-located situations pointing, body language, and using artifacts such as laptops to communicate ``see this'' is easy and affords a low-friction possibility for understanding whether you have caught peoples' attention. In extension, these situations allow the receiving side to see exactly what is meant and allow for clarifying questions to a specific aspect. 
For example, we discussed story-telling aspects of our design. However, discovering the depth and subtlety of the data and how it fits with storytelling concepts appeared overwhelming. Especially as we intended to deepen a shared understanding of this data through visualizations and storytelling, and further, from an interaction perspective, to use scrollytelling to show the complexity of COVID-19 data.
We considered literature on using storytelling elements in visualization~\cite{riche2018data} to prompt meaningful discussions about the data. For example, we referred to the Martini Glass structure~\cite{segel2010narrative} as a potential way to scaffold the complexity of the data and to create cohesion between COVID-19 charts. However, while all team members attempted to grasp these concepts, some team members had a sense of superficially understanding these ideas and found it difficult to use them when thinking about designs.
Despite these challenges, working remotely also poses benefits for knowledge exchange. By working in a distributed team, resources will typically be shared through a medium that allows people to return to them, which is particularly beneficial when acquiring new knowledge.
%\newpage
\challengesection{C7. Understanding Design Ideas} 
%  One of the challenges of collaboration increases the difficulty , it is 
% serendipitous meetings, sharing physical artifacts/sketches
\textit{Working in a distributed design team adds friction to the process of understanding other collaborator’s perspective throughout the design process~\cite{gutwin1998design}. In our case, it took more time to establish understanding when sharing diverse ideas in the process of interpreting domain-specific data and creating data visualizations.}
Typically, a co-located physical space that facilitates the sharing of ideas through face-to-face team activities is used during ideation and iteration phases of the design process. For example, we would compile our sketches on the wall of a meeting room for all members to view and discuss. We adapted this activity to our online synchronous setting by posting sketches to an online collaborative whiteboard in Miro while simultaneously meeting via Zoom. Despite the learning curve of adopting the software, we found this strategy allowed us to share and discuss our ideas productively. Working in the virtual whiteboard environment opened up new opportunities that would not be as feasible when working with the physical counterpart. 
Endless space offered in the virtual whiteboard allowed the team a great deal of flexibility in how the content was laid out. Items on the board could be arranged linearly, and grouped and regrouped as the design process unfolded. The team was able to quickly reference previously shared content, which aided in clarifying how assumptions and misunderstanding had arisen. 
The ability to use virtual pointing and to jump to another's pointing location played a beneficial role in the virtual work space. While pointing is possible in the physical environment, this can at times lack accuracy or require people to physically move closer to the item they are pointing at, which is time-consuming and may occlude it. There is also the limitation that only so many people can be close to a small clipping or sketchbook page. The virtual space permits all collaborators to have an optimal view of where the speaker is pointing and an equal opportunity to point at things.
Iteration was an integral part of the design process, and for this it was often useful to markup existing sketches and design outputs. The digital environment offered the ability to quickly duplicate and markup as many copies of something as needed without compromising the integrity of the original. This was especially enabling when one team member wanted to iterate on another team member's sketch because doing so was immediate and essentially risk-free. %}
   
\section{Strategies for Distributed Design}
   
The intensity of the pandemic situation coupled with the team's general willingness to experiment led us to explore alternative uses of hardware and software. Next, we discuss current potential strategies and potential opportunities that lie ahead. 
\begin{figure*}[tb!]
  \centering
  \includegraphics[width=\linewidth]{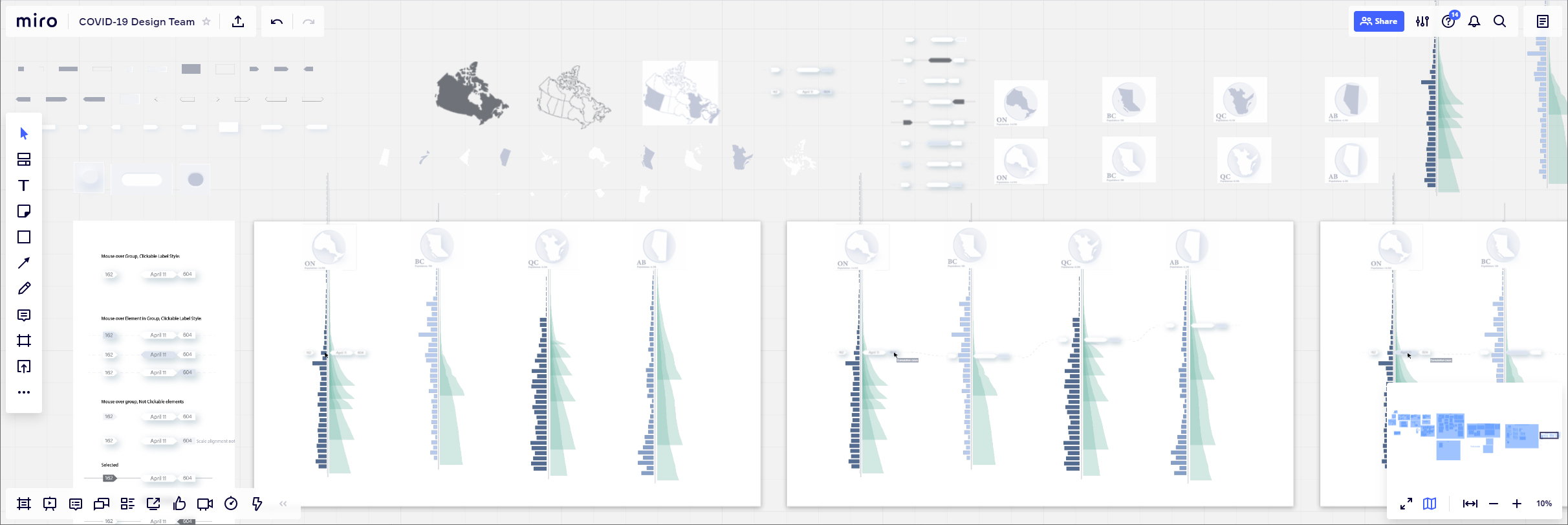}
  \caption{An excerpt of our Miro board that captures a design session for lo-fi prototyping of visualization interaction. During the session all participants were able to duplicate and arrange assets to mimic paper prototyping.}
  \label{fig:prototyping-interactions-in-miro}
\end{figure*}
% \challengesection{C6. Coordinating information about the pandemic}
% This seems irrelevant for the focus of this paper (distributed vis design)
% How the pandemic affected our ability to work. Our experience of working through the pandemic of before and after. Funding, high-level decisions the future of the project.
 
\challengesection{S1. Simulating a Co-located Design Space for Sketching}
\label{subsubsec:simulating-colocation}
\textit {It is difficult to gain a full view of group sketching in a distributed setting because online meetings are generally limited to one view per participant. In an attempt to create a real-time collaborative sketching experience, two team members tested a setup of two device cameras per person during a Zoom meeting.} 
The setup included signing into the meeting twice with two separate devices; one device was a PC camera aimed at the face of each participant and a second device was a phone camera that was directed on their paper and pen. Though it was awkward to find a suitable angle and stabilize the phone, facial expressions, gestures and gaze were captured along with a view of real time sketching. This method allowed the conversation to be held concurrently with a view of sketching practice as it was unfolding within the discussion. This remote sketching activity was seamless in a virtual distributed design setup that simulated a co-located design environment. It supported conversational and visual communication, however, it was cumbersome to recreate this physical setup and it was not attempted after a single trial.
There is considerable potential for developing better and easier ways to set up this type of distributed collaboration. For instance, this approach could bypass knowledge gaps or inaccessibility to using digital sketching tools such as a Wacom tablet. Likewise, we imagine specialised software might provide support for this purpose.
% \begin{figure}[tb]
%   \centering
%   \includegraphics[width=\columnwidth]{figures/sticky-notes-wall-miro2.png}
%   \caption{This is an example of learning in progress. One designer reverted to a physical wall to gather their ideas on stickies, snap a photo and brought it into the Miro board (left). As a group, we wanted the re-positioning freedom of miro board stickies and remade made them to achieve this (right).}
%   \label{fig:sticky-notes}
% \end{figure}
%\challengesection{C12. Testing Design Interactions}
\challengesection{S2. Screen-Sharing to Collaborate in Software Applications}
\textit {Different team members can use their software fluency and bring different skills to the table. With the goal of including everyone throughout the design process, we used screen-sharing via Zoom to collaborate in specialized environments such as designing in Adobe Illustrator or editing code.} 
What made this method particularly useful was when there was skill cross-over in the team because meeting participants could control the mouse during a screen-sharing session. For example, a team member shared their design on the screen of an open Adobe Illustrator workspace during the Zoom meeting. Another team member controlled the mouse from a separate location on their shared Adobe Illustrator workspace. This process allowed for real-time manipulation of visual components of our design and to share skills between multiple team members.
Even in cases where collaborators were not manipulating the software remotely, we found sharing specialized environments to be useful as it mimicked casual co-located collaboration. For example, designers might often gather around the work space of the person implementing a design to tweak position, padding, and other style details. We simulated this experience with screen-sharing and being able to grant remote access to the mouse meant that team members could point out these details with accuracy. 

\challengesection{S3. Using Hand Gestures for Discussing Interaction}
\textit {Animating designs using video or animation software is difficult for team members that lack such technical skills. Not everyone in the group knew how to digitally animate a design, which excluded some team members.}
We discovered that the most accessible way for each team member to show their ideas was by talking camera-to-camera similar to a face-to-face co-located setting. The software knowledge gap was resolved by relying on hand gestures and moving static design artifacts with our hands while remaining visible in the camera. Everyone in the team with access to a web camera could speak to their ideas and show how they imagined the design would move with intuitive hand motions, facial expressions, sounds, pointing to, or moving their own sketches or cut outs while explaining their idea in front of the camera. This emphasized a fuller presence within a remote collaborative experience similar to that of a lab.
We reflected on our choice of communicating about interactions and compared it to our previous visualization design experiences. Based on this, we think we would have pointed to sketches in a co-located mode of collaboration instead of moving our hands in mid-air in front of the camera. We see the hand gestures as adding an extra barrier to communicating about interactions. 
% and assisted showing sketches over teleconferencing. We do note that showing sketches over teleconferencing does not work with virtual backgrounds. % capture diverse ideas. 
% \challengesection{C12. Moving Static Components on a Miro board}
\challengesection{S4. Lo-fi Prototyping for Visualization Interaction}
\textit {It was challenging to find optimal tools to enable all of us to test interactions in a digital format given our variable programming expertise. We aimed to re-create the use of paper prototyping in a digital format.}
To set this up, we used a digital whiteboard in Miro as a ``table top'' space and exported assets from Illustrator to use as ``paper'' clippings. Teleconferencing via Zoom enabled gesturing and speaking into the camera, while all team members could simultaneously copy, resize, and rearrange assets in Miro. Team members were able to collaboratively create UI mock-ups (Figure \ref{fig:prototyping-interactions-in-miro}) and animate them by clicking and dragging. We found this method to be quick and inclusive to all members, regardless of skill. Additionally, we found the ability to duplicate elements and groups of elements without disrupting the integrity of previous models provided a benefit over traditional paper prototyping.

%% file: sections/50-discussion.tex
\section{Discussion}
% Should the discussion be just for the VIS community?
As a team of information visualization designers we described and studied our experience using an auto-ethnographic approach in order to explore how we experienced visualization design methods in our project. Iterative reflection and thematic organization of our experiences presented us with overarching themes of challenges and strategies. During our pursuit to create a vibrant online collaboration in the early stages of visualization design, we were constantly reminded of situations that needed to be addressed, and that we felt were essential to maintain our co-located design process even though we were no longer able to meet in person. In doing so, we identified challenges and strategies that arose through our process, leading us to try new ways to conduct distributed synchronous collaboration. Our account may offer an example to the members of the visualization community who may identify with similar experiences that they too have had. This may serve to inform future visualization research and methodology in this domain.

We found that tools for designing, creating, and using visualizations such as Adobe Illustrator, RAW graphs~\cite{raw}, Tableau, Data Illustrator~\cite{liu2018data}, and Charticulator~\cite{ren2019charticulator} 
did not offer the support we were looking for in our synchronous collaboration. We more frequently relied on external web applications to collaborate both independently and together during our design sessions. We found the combined use of multiple applications such as teleconferencing (in our case, Zoom), virtual whiteboards (in our case, Miro), and collaborative word processors (in our case, Google Docs) provided considerable flexibility. However, constantly switching between applications became cumbersome at times, particularly when coordinating these between several team members during a teleconferencing session. Improved orchestration and integration between these applications, as well as more fine-tuned support for collaborative online sketching would be interesting directions to investigate and welcome improvements in such contexts.

In many design processes there are often activities that require the use of web tools or computing environments and for these we felt the shift to remote work did not hinder our performance of these activities. Research, data understanding, collaborative writing, finessing of high fidelity designs and prototypes are activities that often require the use of a computer. In a co-located scenario, we often have team members gather in a conference room with personal laptops that they must attach to an external display to properly share with the group, or gather at the shoulder of one team member while they informally demo something on their desktop. These types of in-person collaborative moments can be cumbersome. However, we found they happened more rapidly and comfortably over teleconferencing by utilizing built in tools such as screen sharing, or in combination with external collaborative software. 

We struggled more to adapt in the ideation and early iteration phases because our process relied on brainstorming, sketching, and rapid paper prototyping. We tried to mimic real-time collaborative sketching by setting up a dual screen conference, or by holding sketches up to the webcam.  To some extent, we were able to sustain spontaneity of brainstorming and idea sharing similar to those of co-located collaboration by continuing to use mostly physical sketching materials like pen and paper. The transfer of physical sketches to the virtual whiteboard was cumbersome and we see this as an area for potential improvement. However, once the sketches are on a digital whiteboard they continue to be accessible for all future discussions and are easily duplicated and iterated on much more easily than their physical counterparts. 
%We, as designers, needed to think critically about how to tell the story with the data as a way to communicate the complexity of the pandemic in the interest of ensuring that our visualisation would not be misinforming. Our critical discussions often referred to artifacts or relied on gestures and expressiveness that was difficult to convey via teleconference.
While we discussed access to equipment and distraction as being primarily a pandemic related challenge, we also see this as being potentially more widely problematic. Improvements we could make for our sketching sessions might involve high-quality webcams and sketching tablets for all team members. However, the cost and learning curve associated with adopting such technology is high considering that low-fi technology such as pen and paper work very efficiently and with more versatility. Likewise, having dedicated space and time to do one’s work is essential for focus, and offloading the responsibility to the individual to carve it out of their domestic space is understandably challenging for remote workers. 

Where we saw the largest challenge for remote work was communication and social dynamics. Working in a co-located environment facilitates serendipity, natural discussion, and a tangible sense of togetherness that allows for ideas to emerge spontaneously. We found some team members engaged in prolonged teleconferencing sessions to experience the sense of community and to concurrently share ideas as more of a ``hangout'' rather than a meeting. However, other team members who neither could nor chose to utilize video and did not engage in the ``hangouts'' slowly drifted away from the project likely because they felt excluded. In a co-located environment, it is quite clear that an individual is committed to working because they have physically arrived at work, and likewise one can pick up on how busy or idle a co-worker might be when they are co-located. Being disconnected to these physical cues, in conjunction with voluntary roles in our particular project, made assigning tasks or setting expectations challenging. We discovered that social dynamics that seamlessly sort themselves out in a co-located scenario require a lot more facilitation and management effort. We imagine that this may require an additional role or skill set added to a team, or better integration of a person's status (``available'', ``busy'', ``away'') into the collaborative environments. Messaging applications such as Slack provide status information, but in our experience, they lack nuance and integration with realistic highly synchronous workflows. There is no shortage of collaborative software that aims to increase productivity and facilitate focus sessions, but few that fulfill the sense of community and spontaneous collaboration that happens in co-located work environments. 

Finally, it is worth noting that the ability to reflect on our experience of synchronous distributed design is largely due to the fact that we were distributed. We were forced to put every piece of inspiration and every sketch into Miro. All of our notes and communications are documented on Slack and in Google docs. Having this detailed repository at every stage allowed us to reflect and gain insight into our process and learn from our experience.

%% file: sections/60-conclusion.tex
\section{Conclusion}
In this paper, we reflected on our experiences doing distributed visualization design in a remote synchronous design space during the pandemic. We based our reflections on our work on designing a COVID-19 visualization as part of a larger team effort on supporting city- and provincial-level decision-making. We discussed how the pandemic posed new challenges to remote collaboration amidst civic lockdown measures and imposed an increased dependency on spatially distributed teamwork across almost all sectors. As a response to the barriers of working from home being ``the new normal'', we used various synchronous remote design tools and methods with an aim to preserve the richness of co-located collaboration such as face-to-face physical presence with the ability to view real-time body gestures, facial expressions, and the making and sharing of physical artifacts. Based on these, we articulated issues in team composition and communication, both of which affected our visualization design process. We discussed the challenges of working in a distributed visualization design team, such as creating a sense of a shared work environment, which we liken to the idea of a design studio, and how to share data and potential ideas for visualization data. Finally, we offered potential solutions and benefits to working remotely. Our discovery of the challenges and strategies throughout the process provide useful insights about enabling a more fluid distributed collaboration.

While descriptions and discussions of visualization design tended to assume co-located design teams, designers clearly need to carry out some work independently and in different locations. Prior work in other fields have shed light on this. However, there is important work to be done in considering these contexts for visualization design processes. As a step towards developing a better understanding of distributed, synchronous visualization design, this paper provides an experiential understanding and conceptualization of this area.